\documentclass[10pt,aps,prd,preprintnumbers,showpacs,nofootinbib,superscriptaddress,notitlepage]{revtex4-1}

\usepackage{graphicx}
\usepackage{color}
\usepackage{amsmath}
\usepackage{amssymb}
\usepackage[colorlinks=true,citecolor=blue,urlcolor=blue]{hyperref}

\newcommand{\PRE}[1]{{#1}} % Use if preprint style

\newcommand{\cm}{\ensuremath{\mathrm{cm}}}
\newcommand{\kpc}{\ensuremath{\mathrm{kpc}}}
\newcommand{\GeV}{\ensuremath{\mathrm{GeV}}}
\newcommand{\Jtot}{J_\textrm{tot}}

\begin{document}

\title{The Effective \texorpdfstring{$J$}{J}-Factor of the Galactic Center for Velocity-Dependent\texorpdfstring{\\}{} Dark Matter Annihilation}

\author{Kimberly K. Boddy}
\affiliation{Department of Physics \& Astronomy,
Johns Hopkins University, Baltimore, MD 21218, USA}

\author{Jason Kumar}
\affiliation{Department of Physics \& Astronomy,
University of Hawai'i, Honolulu, HI 96822, USA}

\author{Louis E.~Strigari\PRE{\vspace*{.1in}}}
\affiliation{{Department of Physics and Astronomy,
Mitchell Institute for Fundamental Physics and Astronomy,
Texas A\&M University, College Station, TX  77843, USA}}

\begin{abstract}
\PRE{\vspace*{.1in}}
We present the effective $J$-factors for the Milky Way for scenarios in which dark matter annihilation is $p$-wave or $d$-wave suppressed. We find that the velocity suppression of dark matter annihilation can have a sizable effect on the morphology of a potential dark matter annihilation signal in the Galactic Center. The gamma-ray flux from the innermost region of the Galactic Center is in particular suppressed. We find that for dark matter density profiles with steep inner slopes, the morphology of the inner Galaxy gamma-ray emission in $p$-wave models can be made similar to the morphology in standard $s$-wave models. This similarity may suggest that model discrimination between $s$-wave and $p$-wave is challenging, for example, when fitting the Galactic Center excess. However, we show that it is difficult to simultaneously match $s$- and $p$-wave morphologies at both large and small angular scales. The $J$-factors we calculate may be implemented with astrophysical foreground models to self-consistently determine the morphology of the excess with velocity-suppressed dark matter annihilation.
\end{abstract}

\maketitle

%%%%%%%%%%%%%%%%%%%%%%%%%%%%%%%%%%%%%%%%%%%%%%%%%%%%%%%%%%%%%%%%%%%%%%%%%%%%%%%
\section{Introduction}

The Galactic Center (GC) is one of the most interesting targets in searches for gamma rays arising from dark matter annihilation.
This region of sky has been particularly interesting in recent years, as analyses of Fermi-LAT inner Galaxy data have revealed a diffuse gamma-ray emission, which is nearly spherically symmetric about the inner Galaxy, that was not included in previous diffuse models~\cite{Goodenough:2009gk,Hooper:2010mq,Abazajian:2012pn,Daylan:2014rsa}.
This emission is statistically significant, though its precise morphology and energy spectrum is still subject to systematic uncertainties~\cite{Cholis:2012am,Calore:2014xka,Calore:2014nla,TheFermi-LAT:2015kwa,Karwin:2016tsw,TheFermi-LAT:2017vmf} and may still be consistent with an astrophysical population~\cite{Abazajian:2010zy,Bartels:2015aea,Lee:2015fea,Petrovic:2014uda,Cholis:2015dea,Gaggero:2015nsa}.
One way to distinguish signal from background, and to distinguish different types of signals from one another, is to use the morphology of the photon flux.
For the case of standard $s$-wave dark matter annihilation, the annihilation rate scales with the square of dark matter density; thus, for any choice of density profile, one obtains a prediction of the signal flux as a function of the angular distance from the GC.
This morphology is encoded in the $J$-factor, $J(\theta)$, which contains all of the astrophysical dependence of the photon flux.

There are variety of particle physics scenarios in which the dark matter annihilation cross section has a nontrivial dependence on the dark matter particle velocity in the nonrelativistic limit.
This dependence can lead to a suppression or enhancement of the annihilation rate for low-velocity dark matter.
In either case, using the standard $J$-factor is no longer suitable; instead, it is necessary to calculate an effective $J$-factor, which incorporates information from the full dark matter velocity distribution~\cite{Robertson:2009bh,Ferrer:2013cla,Boddy:2017vpe,Bergstrom:2017ptx,Petac:2018gue}.
Since this velocity distribution has a nontrivial spatial dependence, any velocity dependence of the annihilation cross section leads to a departure from the expected angular distribution of the photon flux arising from dark matter annihilation near the GC.

In this paper, we show that the morphology of the GC dark matter signal can change significantly if dark matter annihilation is velocity dependent.
We consider several well-motivated and intuitively simple choices for the dark matter mass distribution near the GC.
In each case, we determine the dark matter velocity distribution using the Eddington formula and then determine the angular dependence of the effective $J$-factor, for the case of either $p$-wave or $d$-wave annihilation.
We find that, as expected, the effective $J$-factor is suppressed for $p$-wave annihilation and is even more suppressed for $d$-wave annihilation.
However, this result is of limited utility, since the overall magnitude of the annihilation cross section also depends on couplings and masses, which in turn depend on the detailed particle physics model.
Of more interest is that, for cuspy profiles, the photon flux arising from $p$-wave dark matter annihilation is less concentrated at small angles than in the case of $s$-wave annihilation.
This effect is even stronger for $d$-wave annihilation.

The plan of this paper is as follows.
In Section~\ref{sec:Jeff}, we describe our method for determining the effective $J$-factor.
In Section~\ref{sec:results}, we present our results, and in Section~\ref{sec:conclusions}, we conclude with a discussion.

%%%%%%%%%%%%%%%%%%%%%%%%%%%%%%%%%%%%%%%%%%%%%%%%%%%%%%%%%%%%%%%%%%%%%%%%%%%%%%%
\section{The Effective \texorpdfstring{$J$}{J}-Factor}
\label{sec:Jeff}

If dark matter is its own antiparticle with an annihilation cross section $\sigma_A$, the resulting differential photon flux produced by annihilation in an astrophysical object can be written as
\begin{equation}
  \frac{d^2\Phi}{dE_\gamma d\Omega} =
  \frac{1}{4\pi} \frac{dN}{dE_\gamma} \int d\ell
  \int d^3 v_1 \frac{f(r(\ell, \theta), \vec{v}_1)}{m_X}
  \int d^3 v_2 \frac{f(r(\ell, \theta), \vec{v}_2)}{m_X}
  \, \frac{(\sigma_A |\vec{v}_1 - \vec{v}_2|) }{2} \ ,
\end{equation}
where $m_X$ is the dark matter particle mass, $dN / dE_\gamma$ is the photon energy spectrum produced by a single annihilation process, and $\ell$ is the distance along the line of sight.
The dark matter particles have velocities $\vec{v}_1$ and $\vec{v}_2$, and the dark matter velocity distribution is given by $f(\vec{r}, \vec{v})$, normalized such that $\int d^3v f(\vec{r}, \vec{v}) = \rho (\vec{r})$, where $\rho$ is the dark matter density.

We parameterize the velocity dependence of the dark matter annihilation cross section, in the nonrelativistic limit, as $\sigma_A v = (\sigma_A v)_0 S(v)$, where $v=|\vec{v}_1 - \vec{v}_2|$ is the relative velocity and $(\sigma_A v)_0$ is the overall amplitude.
We may thus rewrite the differential photon flux as
\begin{equation}
  \frac{d^2 \Phi}{dE_\gamma d\Omega} =
  \frac{(\sigma_A v)_0}{8\pi m_X^2} \frac{dN}{dE_\gamma} J_S (\theta) \ ,
\end{equation}
where
\begin{equation}
  J_S (\theta) =
  \int d\ell \int d^3 v_1 {f(r(\ell, \theta), \vec{v}_1)}
  \int d^3 v_2 {f(r(\ell, \theta), \vec{v}_2)}\, S( |\vec{v}_1 - \vec{v}_2|)
  \label{eq:JSfactor_def}
\end{equation}
is the effective $J$-factor.
In the standard case of $s$-wave annihilation, we have $S(v)=1$, and $J_S(\theta)$ reduces to the more familiar expression $J (\theta) =  \int d\ell \, [\rho (r)]^2 $.

We derive the velocity distribution of dark matter from its density profile using the Eddington formula.
We assume that the velocity distribution is isotropic and spherically symmetric, and thus depends only on $r=|\vec{r}|$ and $v=|\vec{v}|$.
We further assume that the motion of a dark matter particle is determined by an effective smooth potential (\textit{i.e.}, two-body interactions are irrelevant).
These assumptions, along with Jeans's theorem (or, equivalently, along with Liouville's theorem), imply that the time-averaged dark matter velocity distribution depends only the energy, which is the only integral of motion independent of the velocity direction.
Thus, we may write $f(r,v) \equiv f(E(r,v))$, where $E = v^2/2 + \Phi (r) < 0$ is the energy per unit mass, and $\Phi (r) <0$ is the gravitational potential.
Under these assumptions, the velocity and density distributions are related by
\begin{equation}
  \rho(r) = 4\pi \int_0^{v_\textrm{esc}} dv \, v^2 f(r,v)
  = 4\sqrt{2}\pi \int_{\Phi(r)}^0 dE\, f(E) \sqrt{E - \Phi(r)} \ ,
\end{equation}
where the escape velocity $v_\textrm{esc} = \sqrt{-2\Phi(r)}$ depends on the position of a dark matter particle within the halo.
Furthermore, $f(E)$ and $d\rho / d\Phi$ are then related by the Abel integral equation, yielding the Eddington formula
\begin{equation}
  f(E) = \frac{1}{\sqrt{8}\pi^2} \int_E^0
  \frac{d^2 \rho}{d\Phi^2}\frac{d\Phi}{\sqrt{\Phi - E}} \ .
  \label{eq:eddington}
\end{equation}
Given any choice of the density profile $\rho$, one only needs $\Phi (r)$ to determine the velocity distribution.

Under the assumption of spherical symmetry, it is straightforward to obtain the gravitational potential due to dark matter from $\rho (r)$.
The gravitational potential due to baryonic matter near the GC is more complicated, because the observed mass distribution in the bulge and the disk is manifestly nonspherical.
For simplicity and to allow us to utilize the Eddington theory described above, we consider a spherically symmetrized version of the disk and bulge potentials, bearing in mind that the actual distributions are much more complex~\cite{2016ARA&A..54..529B}.
With this simplification, the gravitational potential from the bulge may be approximated by~\cite{Strigari:2009zb,Pato:2012fw}
\begin{equation}
  \Phi_\textrm{bulge}(r) = - \frac{G_N M_b}{r+c_0} \ ,
\end{equation}
where $M_b = 1.5 \times 10^{10} M_\odot$ is the bulge mass and $c_0 = 0.6~\kpc$ is the bulge scale radius.
We model the gravitational potential from the disk by insisting that the mass of the disk contained within a sphere of radius $r$ is the same as the mass contained within the spherical disk potential at radius $r$.
The spherical disk potential is
\begin{equation}
  \Phi_\textrm{disk}(r) = - \frac{G_N M_d}{r} \left[1 - e^{-r/b_d} \right] \ ,
\end{equation}
where $M_d = 7 \times 10^{10} M_\odot$ is the disk mass and $b_d = 4~\kpc$ is the disk scale radius.
To quantify the systematic uncertainty incurred by this approximation for the disk potential, Ref.~\cite{Ferrer:2013cla} compared the predicted dark matter velocity dispersion of a spherical disk potential to that of an axially symmetric disk potential: the velocity dispersion using the spherical disk potential can be up to $\sim 20\%$ larger at intermediate radii, while the velocity dispersions from both potentials agree well at small and large radii.
Models of the baryonic potentials may ultimately be tested with numerical simulations~\cite{Lacroix:2018qqh}.

We thus have all the tools in place for calculating the effective $J$-factor, given a specific form of the dark matter density profile and the velocity dependence of the annihilation cross section.

%%%%%%%%%%%%%%%%%%%%%%%%%%%%%%%%%%%%%%%%%%%%%%%%%%%%%%%%%%%%%%%%%%%%%%%%%%%%%%%
\section{Results}
\label{sec:results}

\begin{figure}[t]
  \centering
  \includegraphics[scale=0.45]{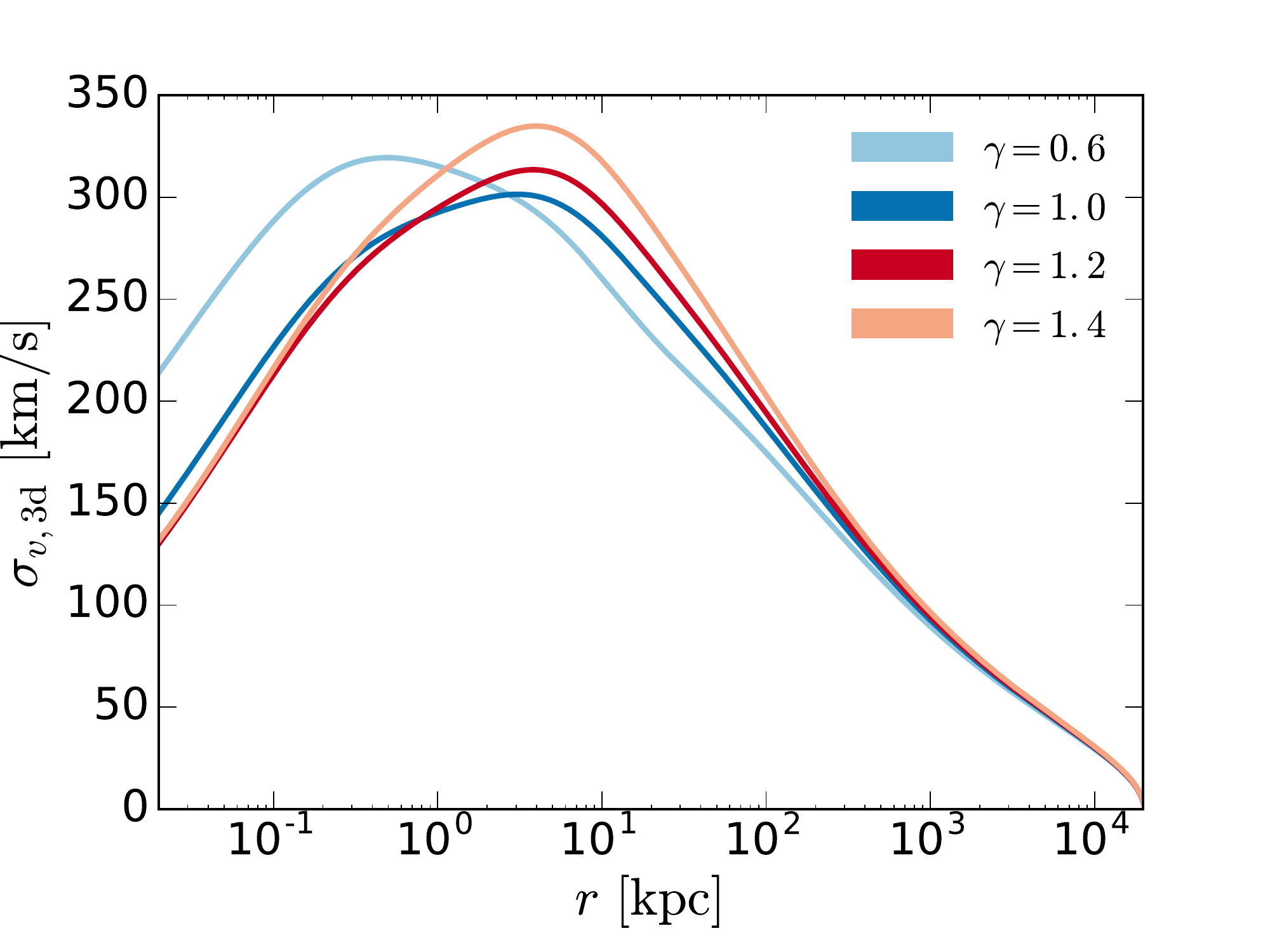}
  \caption{The three-dimensional dark matter velocity dispersion, as a function of distance from the GC.
    We consider modified NFW profiles for benchmark values of the inner slopes $\gamma$, listed in the legend.}
  \label{fig:VelocityDispersion-NFW}
\end{figure}

In addition to the standard scenario of $s$-wave dark matter annihilation with $S(v) = 1$, we also consider $p$-wave annihilation with $S(v) = v^2$ and $d$-wave annihilation with $S(v) = v^4$.
The annihilation of spin-0 or spin-1/2 dark matter to light Standard Model fermions ($XX \rightarrow \bar f f$) from an $s$-wave state is generically suppressed if dark matter is its own antiparticle and if flavor violation is minimal.
In that case, the symmetry properties of the initial state wave function require the $s$-wave ($L=0$) state to have total angular momentum $J=0$; meanwhile, the $\bar f f$ final state particles with $J=0$ must arise from different Standard Model Weyl spinors, implying a matrix element which is suppressed if flavor violation is minimal~\cite{Kumar:2013iva}.
For Majorana fermion dark matter with minimal flavor violation, the dominant contribution to the annihilation cross section may thus be $p$-wave suppressed.
If the dark matter particle is a real scalar, the $p$-wave initial state is antisymmetric under particle exchange and thus forbidden, and the dominant term in the annihilation cross section may instead be $d$-wave suppressed~\cite{Giacchino:2013bta}.

Another well-motivated scenario is Sommerfeld-enhanced dark matter annihilation~\cite{ArkaniHamed:2008qn}, in which case $S(v)$ is large at small $v$.
The effective $J$-factors for this scenario have been considered previously in the context of dwarf spheroidal galaxies (dSphs)~\cite{Boddy:2017vpe,Bergstrom:2017ptx,Petac:2018gue}.
Since the dark matter velocity dispersion at the GC is expected to be much larger than in known dSphs, the enhancement is correspondingly larger in dSphs.
Moreover, the astrophysical foregrounds and backgrounds are under more control for dSphs, making them cleaner targets.
As dSphs are better targets for searching for Sommerfeld enhancements, we do not consider them in this work.
We note, however, that it is possible for coannihilating or multilevel dark matter to have Sommerfeld-enhanced $p$-wave annihilation, in which the annihilation rate in the GC could be higher than that in dSphs~\cite{Das:2016ced}.
Nonetheless, we focus on $p$- and $d$-wave annihilation, for which the photon flux is typically less suppressed at higher relative velocities and thus the GC is the better target.

We consider modified Navarro-Frenk-White (NFW) dark matter density profiles of the form
\begin{equation}
  \rho (r) = \frac{\rho_s}{\left(\frac{r}{r_s}\right)^\gamma
    \left(1 +  \frac{r}{r_s} \right)^{3-\gamma}} \ ,
\end{equation}
where we fix the scale radius to $r_s = 20~\kpc$ and the scale density to $\rho_s = 8 \times 10^6 \, M_\odot /\kpc^3$~\cite{Pato:2015dua,McMillan:2011wd} throughout this work.
We consider the profiles with inner slopes $\gamma = 0.6$, $1.0$, $1.2$, and $1.4$.
For each of these values of the inner slope, the corresponding local dark matter densities at $8.5~\kpc$ are $0.22$, $0.35$, $0.45$, and $0.57~\GeV~\cm^{-3}$, which are consistent with the measured values (see Ref.~\cite{Read:2014qva} and references therein).
In each case, we determine the velocity distribution using Eq.~\eqref{eq:eddington}, including contributions to the gravitational potential from dark matter, the bulge, and the disk.

In Figure~\ref{fig:VelocityDispersion-NFW}, we show the three-dimensional velocity dispersion
\begin{equation}
  \sigma_{v,3\textrm{d}}(r) \equiv
  \frac{\int v^4 f(r,v)\, dv}{\int v^2 f(r,v)\, dv}
\end{equation}
as a function of distance from the GC, for each choice of density profile.
An interesting feature to note is that the dispersion decreases at small distances from the GC, which is due to two effects: the decrease in enclosed mass near the GC and the angular momentum barrier.
With less mass enclosed by a circular orbit close to the GC, the centripetal acceleration is lower, implying a smaller virial velocity.
Moreover, for an isotropic distribution, dark matter particles that are able to reach the very inner part of the GC tend to have a small angular momentum.
The velocity dispersion also decreases and cuts off abruptly at large distances, where the galactic escape velocity is small.
From these general observations, we naively expect $p$- and $d$-wave annihilation to result in photon emission that is suppressed both at small distances from the GC and large distances from the GC.
However, the form of the full dark matter velocity distribution (and thus of the total gravitational potential and dark matter density profile) is necessary to determine the specific behavior of the effective $J$-factor for various annihilation models.

\begin{table}[t]
  \centering
  \begin{tabular}{|l|c|c|c|}
    \hline
    & \multicolumn{3}{|c|}{$\Jtot~[\GeV^2/\cm^5]$} \\
    \cline{2-4}
    $\gamma$ & $s$-wave & $p$-wave & $d$-wave \\
    \hline
    0.6 & 3.27e+22 & 5.58e+16 & 1.86e+11 \\
    1.0 & 1.30e+23 & 2.42e+17 & 9.53e+11 \\
    1.2 & 3.21e+23 & 6.07e+17 & 2.62e+12 \\
    1.4 & 1.05e+24 & 1.93e+18 & 9.13e+12 \\
    \hline
  \end{tabular}
  \caption{Table of the total integrated effective $J$-factor, $\Jtot \equiv \int d\Omega \, J_S (\theta)$.
    We consider modified NFW profiles, with a scale radius of $20~\kpc$ and scale density of $8 \times 10^6 \, M_\odot /\kpc^3$, for benchmark values of the inner slopes $\gamma$.}
  \label{tab:Jtot}
\end{table}

\begin{figure}[t]
  \centering
  \includegraphics[width=0.48\linewidth]{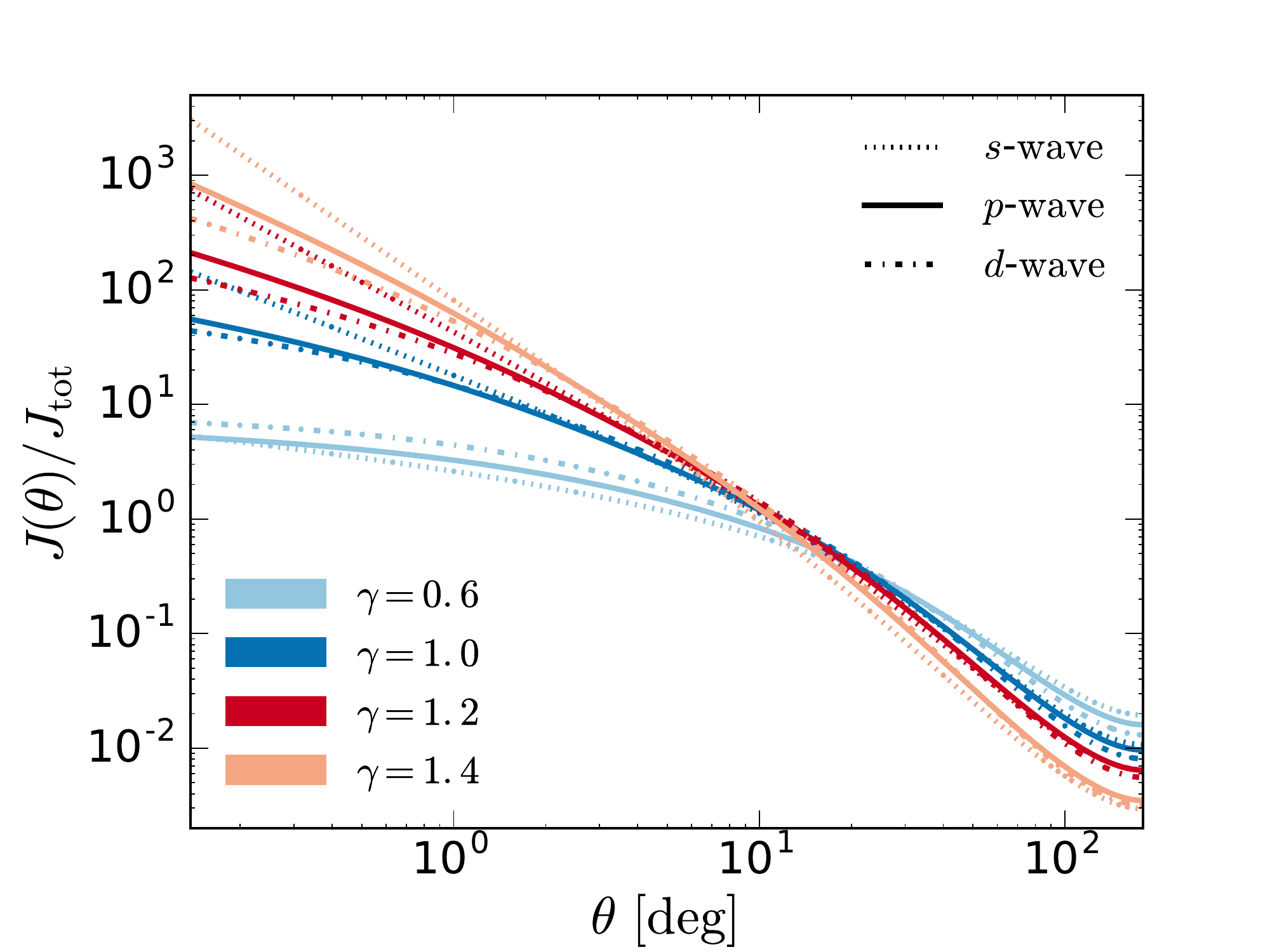}
  \hspace{0.2in}
  \includegraphics[width=0.48\linewidth]{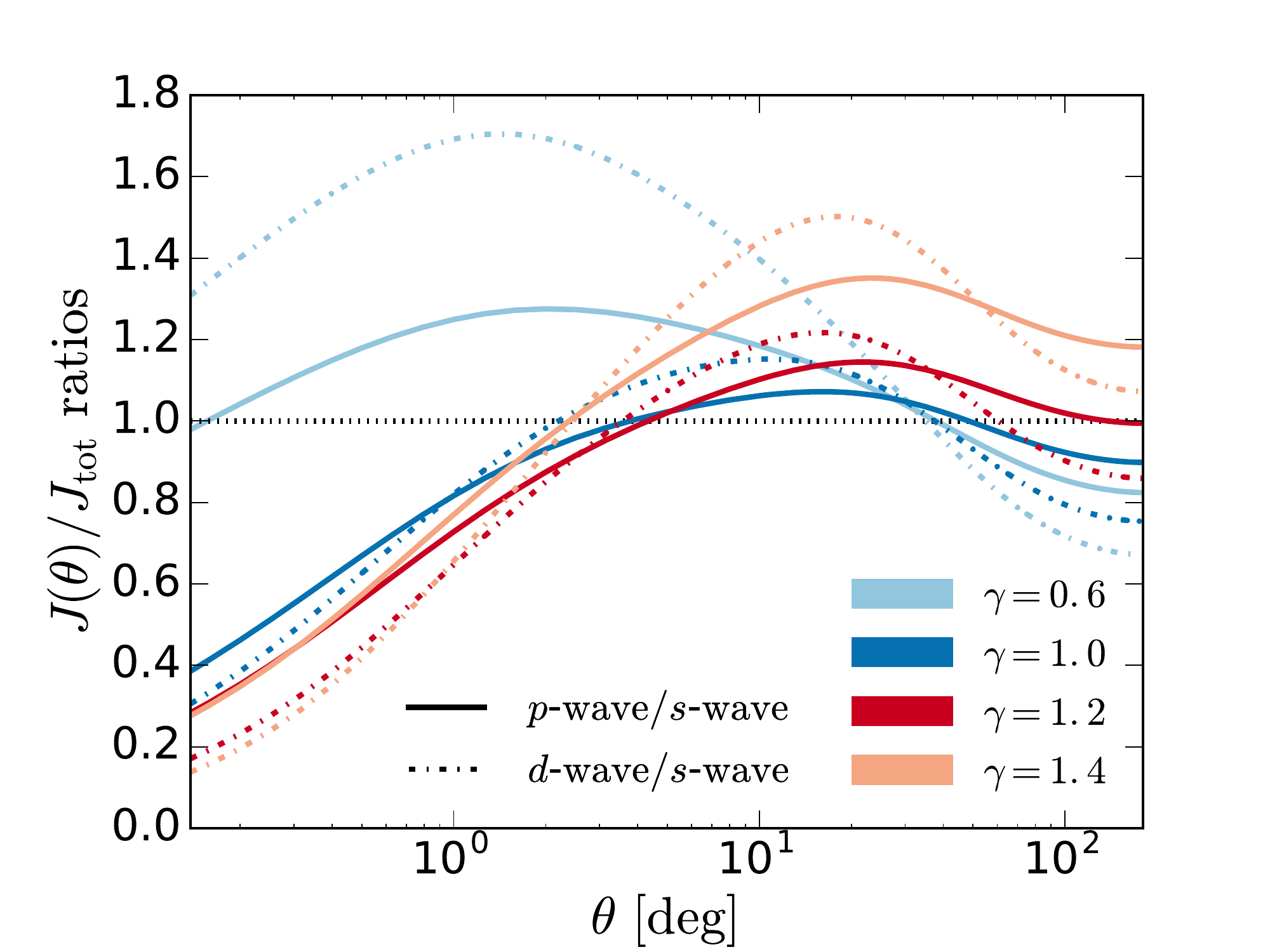}
  \caption{\textbf{[Left]:} The normalized angular distribution, as a function of angle from the GC, of the photon flux arising from $s$-, $p$-, and $d$-wave annihilation.
    We consider modified NFW profiles for benchmark values of the inner slopes $\gamma$, listed in the legend.
    \textbf{[Right]:} Similar to the left panel, except we show the ratios of the normalized angular distributions for $p$-wave to $s$-wave annihilation and for $d$-wave to $s$-wave annihilation.
    We show a reference line (dotted black) at 1, indicating where the distributions of the different annihilation models coincide.}
  \label{fig:Jfactors-NFW}
\end{figure}

We calculate the effective $J$-factors for our benchmark values of the inner slope $\gamma$, for $s$-, $p$-, and $d$-wave annihilation.
For each case, we normalize $J_S(\theta)$ by the total effective $J$-factor, $\Jtot \equiv \int d\Omega \, J_S (\theta)$, integrated over all angles.
We report the values of $\Jtot$ in Table~\ref{tab:Jtot}.
In the left panel of Figure~\ref{fig:Jfactors-NFW}, we show the quantity $J_S(\theta)/\Jtot$, which is the angular distribution of the photon flux.
Since the benchmark dark matter profiles have the same scale radius and scale density, the values of $J_S(\theta)$ are larger for steeper inner slopes (larger values of $\gamma$).
In particular, differences in $J_S(\theta)$ for various $\gamma$ are much larger at small angles, where more of the dark matter is concentrated; away from the GC, the benchmark profiles have the same outer slopes, so their line-of-sight integrations at large angles are somewhat similar (within an order of magnitude).
Note that the angular distribution at large angles is more suppressed for larger values of $\gamma$, due to the larger normalization factor $\Jtot$.

In the right panel of Figure~\ref{fig:Jfactors-NFW}, we compare the various annihilation models for a given choice of the dark matter density profile; we show the ratios of the angular distributions for $p$- and $d$-wave annihilation to that for $s$-wave annihilation.
For a pure NFW profile, within the inner $1^\circ$ of the GC and at large angles $\gtrsim 50^\circ$, the angular distribution is suppressed for the case of $p$- or $d$-wave annihilation, relative to $s$-wave annihilation.
In the range $2-50^\circ$, the distribution for $p$- and $d$-wave annihilation is enhanced.
These results are consistent with our expectations, given the behavior of the dark matter velocity distribution.
Comparing the relative suppression/enhancement from $p$- and $d$-wave annihilation for different values of $\gamma$ requires more care.
For cuspier profiles, velocity-suppressed annihilation yields a more heavily suppressed angular distribution in the inner $1^\circ$, as more of the dark matter is concentrated at very small distances from the GC, where the relative velocities are small.
For less cuspy profiles, however, there is less dark matter at small distances from the GC, and the relative velocities are slightly higher.
As a result, the effect of velocity suppression near the GC is less severe, while the effect far from the GC is more important.
For the case of $\gamma =0.6$, we indeed find that velocity-suppressed annihilation leads to an enhancement in the angular distribution within the inner $1^\circ$ and a suppression at large angles, compared to the case of a pure NFW profile.

We expect the generic features of velocity-suppressed annihilation to hold for more general dark matter profiles and even for different models of the baryonic potential.
For an isotropic dark matter distribution, the suppression of the velocity dispersion both close to and very far from the GC is a generic feature resulting from the angular momentum barrier and the behavior of the virial and escape velocities, regardless of the particular choice of dark matter profile or baryonic potential.
Moreover, as previously discussed, modeling the disk with either a spherical or an axially symmetric potential yields similar dark matter velocity dispersions at small and large radii~\cite{Ferrer:2013cla}.
We expect an axially symmetric disk to have little impact on our results at small and large angles, apart from an overall reduction of $\Jtot$.
Numerical simulations may be used to test this assumption for the potential and the assumption of isotropy at different radii; simulations currently find a slight deviation from isotropy near the solar radius~\cite{Ling:2009eh,Bozorgnia:2016ogo}.

%%%%%%%%%%%%%%%%%%%%%%%%%%%%%%%%%%%%%%%%%%%%%%%%%%%%%%%%%%%%%%%%%%%%%%%%%%%%%%%
\section{Conclusions}
\label{sec:conclusions}

We have computed the effective $J$-factors for the GC, relevant for velocity-dependent dark matter annihilation.
In particular, we focus on the well-motivated cases of $p$-wave and $d$-wave suppressed annihilation into photons.
We find that properly incorporating the velocity dependence of the annihilation can noticeably affect the morphology of an annihilation signal.
For models with velocity-suppressed annihilation, the photon angular distribution is enhanced at angles of ${\cal O}(10^\circ)$ from the GC, but is suppressed within the inner $1^\circ$ if the profile is cuspy.
The suppression of the photon flux at small angles is a result of angular momentum conservation: due to the presence of an angular momentum barrier, the particles that are able reach the innermost part of the GC tend to have lower kinetic energies.
This effect is more pronounced for cuspy profiles, in which more dark matter is concentrated near the GC.
For less cuspy profiles, the more important effect is the suppression of the velocity at large distances, which in turn suppresses the angular distribution at large angles.

We have assumed a modified NFW profile, modeling the gravitational potential due to baryons as spherically symmetric contributions from the bulge and the disk.
A variety of other astrophysical models are also possible.
We expect that the qualitative behavior of the angular distribution for velocity-suppressed annihilation to be robust against these systematic uncertainties in the dark matter and baryonic matter distributions, as the angular momentum barrier, the decrease of the enclosed mass at small radii, and the decrease in the escape velocity with distance from the GC are universal features.
However, determining whether or not the velocity-suppressed annihilation leads to a suppression or an enhancement of the angular distribution at various angles---compared to that from the standard case of velocity-independent annihilation---does depend on the form of the dark matter and baryonic matter distributions and requires a full calculation.

It is interesting to apply these results to the excess of ${\cal O}(\GeV)$ photons from the GC seen in Fermi-LAT data~\cite{Goodenough:2009gk,TheFermi-LAT:2015kwa}.
If this excess arises from velocity-suppressed annihilation of dark matter with a cuspy profile, it should be suppressed within the innermost $1^\circ$ of the GC.
Fermi-LAT data from the innermost $1^\circ$ (which is the approximate angular resolution of the Fermi-LAT) can thus be used to test scenarios in which the origin of the GC excess is velocity suppressed dark matter annihilation.
In Table~\ref{tab:1deg}, we provide the integrated angular distributions to find the fraction of photon emission arising from a cone of radius $1^\circ$ about the GC.

For cuspy profiles, the fraction of the photon flux emanating from the inner $1^\circ$ can be suppressed by up to a factor of 2 as a result of velocity-suppressed annihilation.
It is possible to counteract this suppression with a steeper inner slope.
For the case of $\gamma =1.2$ and $s$-wave annihilation, about 10\% of the flux arises from the inner $1^\circ$; while for the case of $\gamma=1.4$ and $p$-wave annihilation, about 14\% of the flux arises from the inner $1^\circ$.
But as we see from Figure~\ref{fig:Jfactors-NFW}, although choosing a steeper profile can compensate for the effects of velocity suppression within the inner $1^\circ$, they also cause a dramatic suppression of the angular distribution for $\theta \gtrsim {\cal O}(10^\circ)$.
Changing the inner slope alone can mask the effects of velocity-suppressed annihilation on the signal morphology at small angles or at large angles, but not both.

Of course, any analysis of the morphology of the GC excess necessarily relies on assumptions about the morphology of astrophysical foregrounds and backgrounds in the direction of the GC.
A detailed study the consistency of the morphology of the GC excess with velocity-suppressed dark matter annihilation, and of the potential of future instruments with better angular resolution, is an interesting topic that we leave for future work.

\begin{table}[t]
  \centering
  \begin{tabular}{|l|c|c|c|}
    \hline
    $\gamma$ & $s$-wave & $p$-wave & $d$-wave \\
    \hline
    0.6 & 2.97e-03 & 3.54e-03 & 4.80e-03 \\
    1.0 & 3.11e-02 & 2.06e-02 & 1.94e-02 \\
    1.2 & 1.00e-01 & 5.33e-02 & 4.21e-02 \\
    1.4 & 2.68e-01 & 1.37e-01 & 9.75e-02 \\
    \hline
  \end{tabular}
  \caption{Table of the fraction of the photon flux arising from a $1^\circ$ degree cone about the GC, assuming $s$-, $p$-, or $d$-wave annihilation.
    We consider modified NFW profiles, with a scale radius of $20~\kpc$ and scale density of $8 \times 10^6 \, M_\odot /\kpc^3$, for benchmark values of the inner slopes $\gamma$.}
  \label{tab:1deg}
\end{table}

%%%%%%%%%%%%%%%%%%%%%%%%%%%%%%%%%%%%%%%%%%%%%%%%%%%%%%%%%%%%%%%%%%%%%%%%%%%%%%%
\vskip 0.1in
{\bf Acknowledgements.}
We are grateful to Alex Drlica-Wagner, Bhaskar Dutta, Andrew Pace, and Aaron Vincent for useful discussions and feedback.
KB acknowledges KITP and {\it The Small-Scale Structure of Cold(?) Dark Matter} workshop for their hospitality and support under NSF grant \#PHY-1748958, during the completion of this work.
The work of JK is supported in part by NSF CAREER Grant No.~PHY-1250573.
The work of LES is supported by DOE Grant de-sc0010813.

%%%%%%%%%%%%%%%%%%%%%%%%%%%%%%%%%%%%%%%%%%%%%%%%%%%%%%%%%%%%%%%%%%%%%%%%%%%%%%%

%%%%%%%%%%%%%%%%%%%%%%%%%%%%%%%%%%%%%%%%%%%%%%%%%%%%%%%%%%%%%%%%%%%%%%%%%%%%%%%

\end{document}